\documentstyle[prd,aps,twocolumn,floats,epsfig,amsmath2000,amsfonts]{revtex}

\DeclareMathOperator{\diag}{{\rm diag}}

\newcommand{\Dsl}{\rlap{\,/}{{D}}}

\newcommand{\RMT}{{\rm RMT}}
\newcommand{\disc}{{\rm disc}}
\newcommand{\conn}{{\rm conn}}
\newcommand{\chidisc}{\chi^{\disc}}
\newcommand{\chiconn}{\chi^{\conn}}
\newcommand{\abs}[1]{\lvert#1\rvert}
\newcommand{\dsum}{\displaystyle\sum}

\newcommand{\SU}{{\rm SU}}
\newcommand{\U}{{\rm U}}
\newcommand{\setR}{\mathbb{R}}

\newcommand{\cc}[1]{\multicolumn{1}{c}{#1}}
\newcommand{\Cc}[1]{\multicolumn{2}{c}{#1}}

\begin{document}
\fussy
\draft
\wideabs{
\title{Spectrum of the SU(3) Dirac operator on the lattice:\\
       Transition from random matrix theory to chiral perturbation
       theory}
\author{M.~G{\"o}ckeler, H.~Hehl, P.~E.~L.~Rakow, A.~Sch{\"a}fer}
\address{Institut f{\"u}r Theoretische Physik,
  Universit{\"a}t Regensburg, D-93040 Regensburg, Germany}
\author{T.~Wettig}
\address{Department of Physics, Yale University, New Haven, CT
  06520-8120, USA\\ RIKEN-BNL Research Center, Brookhaven National
  Laboratory, Upton, NY 11973-5000, USA}
\date{May 14, 2000}    
\maketitle
\begin{abstract}
  We calculate complete spectra of the Kogut--Susskind Dirac operator
  on the lattice in quenched SU(3) gauge theory for various values of
  coupling constant and lattice size.  From these spectra we compute
  the connected and disconnected scalar susceptibilities and find
  agreement with chiral random matrix theory up to a certain energy
  scale, the Thouless energy.  The dependence of this scale on the
  lattice volume is analyzed.  In the case of the connected
  susceptibility this dependence is anomalous, and we explain the
  reason for this.  We present a model of chiral perturbation theory
  that is capable of describing the data beyond the Thouless energy
  and that has a common range of applicability with chiral random
  matrix theory.
\end{abstract}
}

\narrowtext

\section{Introduction}

It has by now been well established that the low-lying eigenvalues of
the lattice QCD Dirac operator can be described by chiral random
matrix theory (chRMT) \cite{ShuVer} if the linear size $L$ of a
lattice with Euclidean 4-volume $V=L^4$ fulfills the constraints
\begin{equation}
  \label{domain}
  1/\Lambda \ll L \ll 1/m_\pi \:,
\end{equation}\\[-0.8ex]
where $\Lambda$ is a typical hadronic scale such as the rho mass, and
$m_\pi$ is the pion mass\cite{review1,review2}.  The first inequality
means that the low-energy features of the theory can be described in
terms of Goldstone modes governed by an effective chiral Lagrangian,
whereas the second inequality tells us that the zero-momentum modes
make the dominant contribution to the partition function so that the
kinetic terms in the chiral Lagrangian can be neglected.  This implies
that in the phase where chiral symmetry is spontaneously broken the
low-lying eigenvalues of the Dirac operator are only sensitive to
symmetry properties of the underlying theory.  For staggered fermions
one finds that the chiral Gaussian symplectic ensemble (chGSE) of
chRMT corresponds to $\SU(2)$ gauge theory with fermions in the
fundamental representation, the chiral Gaussian orthogonal ensemble
(chGOE) is relevant for $\SU(N_c)$, $N_c\geq2$, gauge theory with
adjoint fermions, and the chiral Gaussian unitary ensemble (chGUE)
describes $\SU(N_c)$, $N_c\geq3$, gauge theory with fundamental
fermions \cite{HV,EdwHel}.  Agreement with chRMT has been confirmed
for all three ensembles, primarily for the microscopic spectral
density of the Dirac operator
\cite{EdwHel,Berb98a,Ma98,Berb98b,SU3,our}.

In earlier papers \cite{Jac96,Jac98,Zahed98,Ber2} the energy scale up
to which chRMT applies, the so-called Thouless energy, has been
predicted theoretically and verified in $\SU(2)$ lattice simulations.
In these investigations it has been found that this upper energy scale
$\lambda_{\RMT}$ is given by
\begin{equation}
  \label{Thouless}
  \lambda_{\RMT}/\Delta \sim f_\pi^2 L^2 \, ,
\end{equation}
where $\Delta=1/\rho(0)=\pi/(V\Sigma)$ is the mean level spacing at
virtuality zero as given by the Banks--Casher formula. Here, $f_\pi$
is the pion decay constant (normalized such that $f_\pi=93\:{\rm MeV}$
in the real world), $\Sigma$ is the absolute value of the 4-flavor
chiral condensate for infinite volume and vanishing mass, and
$\rho(\lambda)$ is the spectral density of the Dirac operator averaged
over gauge field configurations.

In this paper we extend our analysis of the Thouless energy to the
physically more interesting case of $\SU(3)$. We concentrate on the
chiral susceptibilities $\chidisc$ and $\chiconn$ which in the
$\SU(2)$ case were shown to be especially well suited for such an
analysis \cite{Ber2,ourSU2}.  For an earlier analysis of $\chidisc$ in
the $\SU(3)$ case and a general comparison of chRMT to our data we
refer to Ref.~\cite{our}.  In order to describe the Dirac spectrum
beyond the Thouless energy, we then construct a model of chiral
perturbation theory that takes into account the differences between
Kogut--Susskind fermions at finite lattice spacing and continuum
fermions.  A similar analysis using continuum chiral perturbation
theory was performed in the framework of an instanton liquid model in
Ref.~\cite{Jac98}.

\section{Simulations}

The Euclidean Kogut--Susskind Dirac operator on the lattice reads
\begin{equation}
  \label{DiracOp}
  \Dsl_{x,y} =
       \frac12\sum_\mu\bigl[\eta_\mu(x)U_\mu(x)\delta_{x+\hat\mu,y}
       -\eta_\mu(y)U^{\dag}_\mu(y)\delta_{x-\hat\mu,y}\bigr] \,,
\end{equation}
where $U$ and $\eta$ denote the link variables and the staggered
phases, respectively, and the lattice spacing has been set equal to
one, i.e., we use lattice units.  The Dirac operator in
Eq.~\eqref{DiracOp} is anti-hermitian.  Hence its eigenvalues ${\rm
  i}\lambda_k$ are purely imaginary ($\lambda_k\in\setR$).
Furthermore, the spectrum is symmetric about zero, i.e., for each
eigenvalue $\lambda_k\neq0$ there is another eigenvalue $-\lambda_k$.
Note that for $N_c\geq3$ the eigenvalues are generically
non-degenerate, other than in $\SU(2)$ where each eigenvalue is doubly
degenerate.  We compute spectra of $\Dsl$ using the same methods as in
Ref.~\cite{our}.
\begin{table}
  \begin{tabular}{dcccc}
    $\beta\,\backslash V$ &  $4^4$  &  $6^4$  &  $8^4$  &  $10^4$   \\
    \hline \\[-2mm]
    $5$.$2$    & $40830$ & $24708$ &  $7665$ &  $5806$   \\
    $5$.$4$    & $35337$ & $24210$ &  $6000$ &  $6300$   \\
    $5$.$6$    & $36158$ & $21000$ &  $8000$ &  $7700$   \\
    $5$.$7$    &         & $19000$ &  $6000$ &  $6198$
  \end{tabular}
  \vspace*{2mm}
  \caption{Number of generated configurations.}
  \label{configs}
\end{table}

\begin{table}
  \begin{tabular}{dllll}
    $\beta\,\backslash
                    V$&\cc{$4^4$}&\cc{$6^4$}&\cc{$8^4$}&\cc{$10^4$}\\
    \hline \\[-2mm]
    $5$.$2$ & $1$.$08(4)$ & $1$.$06(3)$ & $1$.$06(4)$ & $1$.$07(4)$\\
    $5$.$4$ &$0$.$849(30)$&$0$.$875(20)$& $0$.$86(2)$ & $0$.$87(2)$\\
    $5$.$6$ & $0$.$47(10)$& $0$.$47(2) $& $0$.$47(2)$ & $0$.$47(3)$\\
    $5$.$7$ &  \cc{}      &$0$.$255(80)$&$0$.$242(40)$&$0$.$255(50)$
  \end{tabular}
  \vspace*{2mm}
  \caption{Absolute values of the chiral condensate $\Sigma$.}
  \label{Sigmas}
\end{table}\nopagebreak
Table~\ref{configs} gives the number of configurations for the values
of $\beta=6/g^2$ (with $g$ the coupling constant) and the lattice
volume $L^4$ that we used.  The absolute values of the
(unrenormalized) chiral condensate $\Sigma$ together with the
statistical errors are given in Table~\ref{Sigmas}.  They were
computed from a fit of the distribution of the smallest positive
eigenvalue $\lambda_{\rm min}$ to the chRMT formula \cite{Forrester}
$P(\lambda_{\rm min})=\frac12(V\Sigma)^2\lambda_{\rm min}\, {\rm
  e}^{-(V\Sigma\lambda_{\rm min})^2/4}$.

\section{Scalar susceptibilities}

Consider a lattice theory with two kinds of quarks, a valence quark
with mass $m_v$ and a sea quark with mass $m_s$. We will take $N_v$
generations of valence quarks and $N_s$ generations of sea quarks.
Each generation corresponds to $4$ flavors in the continuum limit.
The partition function $Z$ for the theory is given as a sum over gauge
field configurations $U$,
\begin{equation}
   Z(m_v,m_s)= \sum_U{\rm e}^{-S_g(U)}
          \det(m_v + \Dsl)^{N_v} \det(m_s + \Dsl)^{N_s},
\end{equation}\\
where $S_g(U)$ is the gauge action.  The definitions of the chiral
condensate $\sigma$ and the two scalar susceptibilities for $N_c\geq3$
are then
\begin{equation}
  \sigma(m_v,m_s) = \lim_{N_v\to0}
  \dfrac1{VN_v}\dfrac\partial{\partial m_v}\ln
        Z(m_v,m_s) \;, \label{a}
\end{equation}
\begin{align}
  \chiconn(m_v) &= \left.\dfrac\partial{\partial m_v}\sigma(m_v,m_s)
                   \right|_{m_s=m_v} ,\\
  \chidisc(m_v) &= \dfrac1{N_s}\left.\frac\partial{\partial
                  m_s}\sigma(m_v,m_s)\right|_{m_s=m_v} \label{c} .
\end{align}

For quenched ($N_s=0$) Kogut--Susskind fermions $\sigma$ depends only
on the valence quark mass $m_v=m$.  From the complete spectra of
$\Dsl$ we compute $\sigma$ for arbitrary values of $m$ according to
\begin{equation}
 \sigma_{\rm lattice}(m) = \frac{1}{V} \left\langle
 \sum_{k=1}^N \frac{1}{{\rm i}\lambda_k+m} \right\rangle\:,
\end{equation}
where $N=N_cV$ is the number of the eigenvalues ${\rm i}\lambda_k$ of
$\Dsl$ and the average is over gauge field configurations.  The
susceptibilities are given by
\begin{align}
 \chi^{\rm conn}_{\rm lattice}(m)
  ={}& \frac{\partial}{\partial m} \sigma(m)
  = -\frac1V\left\langle\sum_{k=1}^N\frac1{({\rm i}\lambda_k+m)^2}
                    \right\rangle
\intertext{and}
\begin{split}
  \chi^{\rm disc}_{\rm lattice}(m)
  ={}& \dfrac{1}{V}\left\langle\dsum_{k,\ell=1}^N
    \dfrac{1}{({\rm i}\lambda_k+m)({\rm i}\lambda_\ell+m)}
      \right\rangle \\
  &-\dfrac{1}{V} \left\langle\dsum_{k=1}^N\dfrac{1}{{\rm
        i}\lambda_k+m}\right\rangle^2 \;.
\end{split}
\label{e4}
\end{align}
In the last equation the limit $m_s=m_v=m$ has been taken after
evaluating the derivative in Eq.~\eqref{c}.  Note that slightly
different definitions are used for gauge group $\SU(2)$ because of the
degeneracy of the eigenvalues, see Ref.~\cite{ourSU2}.

From chRMT one obtains expressions for the chiral condensate and the
susceptibilities which depend on $N_s$ and on the topological charge
$\nu$.  In order to compare with our numerical data we have to set
$N_s=0$ because we work in the quenched approximation.  We shall also
set $\nu=0$, since the staggered Dirac operator has no exact zero
modes (not even approximate ones because of the relatively strong
couplings we use) so that we are effectively in the sector of
vanishing topological charge.  Thus we get from the chGUE, which is
the appropriate ensemble for the gauge group $\SU(3)$, the following
result for the chiral condensate,
\begin{equation}
  \frac{\sigma_{\RMT}}{\Sigma} = u\bigl[I_0(u)K_0(u) +
  I_1(u)K_1(u)\bigr]\;,
\end{equation}
where the rescaled mass parameter $u$ is given by
\begin{equation}
  u=m\Sigma L^4\:.
\end{equation}
The functions $I_n, K_n$ are modified Bessel functions. For the
connected susceptibility one has
\begin{align}
  \frac{\chiconn_{\RMT}}{V\Sigma^2} &= I_0(u)K_0(u) - I_1(u)K_1(u)\;,
\intertext{and for the disconnected susceptibility}
  \frac{\chidisc_{\RMT}}{V\Sigma^2} &=
    u^2\bigl[I_0^2(u)-I_1^2(u)\bigr]\bigl[K_1^2(u)-K_0^2(u)\bigr]\;.
\end{align}
Once $\Sigma$ has been fixed, the chRMT predictions do not contain any
free parameters.

\section{Scaling of the Thouless energy}

\begin{figure}[htb]
  \begin{center}
   \epsfig{file=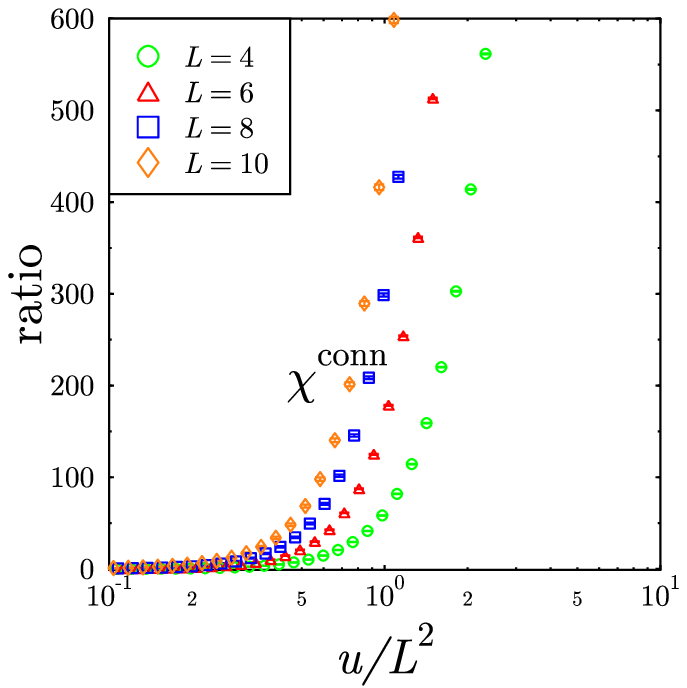,width=77mm}\\[2mm]
   \epsfig{file=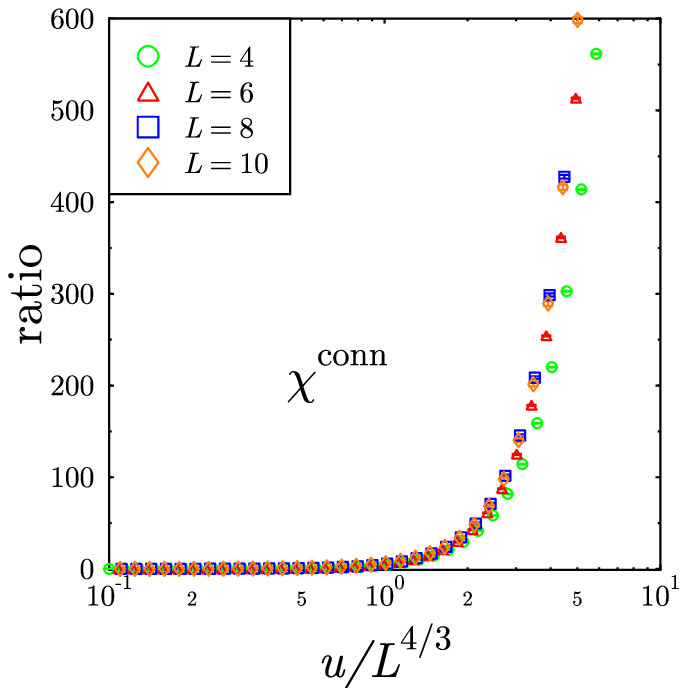,width=77mm}\\[4mm]
    \caption{Ratios $(\chiconn_{{\rm lattice}}-\chiconn_{\RMT})/
      \chiconn_{\RMT}$ of the connected scalar susceptibility
      $\chiconn$ for SU(3) staggered fermions at $\beta=5.2$ and
      various lattice volumes $V=L^4$.}
    \label{ratios}
  \end{center}
\end{figure}

In Refs.~\cite{Ber2,ourSU2,beyond} it was demonstrated for the gauge
group $\SU(2)$ that chRMT describes $\chiconn$ and $\chidisc$
perfectly up to a value of $u$ which scales like $L^2$, in agreement
with Eq.~\eqref{Thouless}.  The same scaling was found for $\chidisc$
in the case of $\SU(3)$ \cite{our}.  Interestingly, this is not true
for $\chiconn$ in $\SU(3)$ \cite{beyond}, where we find instead a
scaling with $L^{4/3}$, see Fig.~\ref{ratios}.
This surprising behavior can be traced back to the form of the
asymptotic expansion of $\chiconn_{\RMT}$ as we shall now explain.

Let us assume that the susceptibilities can be described by
\begin{equation}
  \chi \approx \chi_{\rm RMT} + {\rm constant}\:.
\end{equation}
(We will present a more sophisticated model in terms of chiral
perturbation theory later in this article.) The constant describes the
large-mass limit ($u\to\infty$) of $\chi$ that will be of order $1$ in
lattice units.  In this limit finite-size effects are negligible, and
the constant is therefore also the thermodynamic limit of $\chi$.  It
will become important when the mass has become so large that
$\chi_{\RMT}$ has dropped to values of order $1$.

Now, the asymptotic behavior ($u\to\infty$) of the chRMT predictions
of the two susceptibilities depends on the color group and is given in
Table~\ref{asymp}. Note the different powers of $u$ in the denominator
of $\chiconn_{\RMT}$.
\begin{table}[h]
  \begin{tabular}{ccc}
    &   SU(2) -- chGSE   &   SU(3) -- chGUE\\
    \hline\\[-2mm]
    $\chidisc_{\RMT}$ & $\dfrac{V\Sigma^2}{8u^2}+{\cal
      O}(u^{-3})$& $\dfrac{V\Sigma^2}{4u^2}+{\cal O}(u^{-4})$\\
    \\
    $\chiconn_{\RMT}$ & $\dfrac{V\Sigma^2}{4u^2}+{\cal
      O}(u^{-3})$& $\dfrac{V\Sigma^2}{4u^3}+{\cal O}(u^{-4})$\\[3mm]
  \end{tabular}
  \vspace*{2mm}
  \caption{Asymptotic expansions of the two susceptibilities with
    different gauge groups.}
  \label{asymp}
\end{table}

In the standard case, where the expansion starts with $1/u^2$, $\chi$
gets of order $1$ at $u^2\sim V\Sigma^2$, i.e., $u \sim L^2$.  This is
the case of $\SU(2)$ and has a natural explanation, namely, that the
Compton wavelength of the pion becomes comparable with the box size
$L$ of the lattice. This follows from the upper bound of
\eqref{domain} and Eq.~\eqref{Thouless}.  In the case of $\SU(3)$
$\chiconn\sim1$ when $u^3\sim V\Sigma^2$ and so the value of $u$ where
RMT breaks down scales with $L^{4/3}$ instead.

Where does the unusual power come from? It is a consequence of the
quenched formulation and of the fact that our lattice simulations are
effectively in the sector of topological quantum number $\nu=0$. The
general chRMT prediction for arbitrary topological quantum number
$\nu$ and number of flavors $N_s$ is given by
\begin{equation}
  \label{genchiconn}
    \frac{\chiconn_{\RMT}}{V\Sigma^2} = I_n(u)K_n(u) -
    I_{n+1}(u)K_{n-1}(u)
\end{equation}
with $n=N_s+\abs{\nu}$. This leads to an asymptotic expansion
\begin{equation}
  \frac{\chiconn_{\RMT}}{V\Sigma^2}  \sim \frac n{u^2} -
  \frac{4n^2-1}{4u^3} +{\cal O}(u^{-4}) \:.
 \label{lead_term}
\end{equation}
Thus, unless $N_s=\nu=0$ (as in our case) one should recover the usual
behavior.  It would be very desirable to check this prediction in
unquenched lattice simulations ($N_s\neq0$) or with Ginsparg--Wilson
fermions, which can reproduce the $\nu\neq0$ sectors of QCD
\cite{topology}.

\section{Chiral perturbation theory}

We want to describe the lattice data also beyond the Thouless energy
and therefore have to use a physical model that goes beyond RMT.  One
description of the low-energy limit of QCD is chiral perturbation
theory (chPT).  Since all our data are from quenched simulations we
actually have to use the quenched version of chPT.

The data we consider are rather far from the continuum limit, so we
cannot rely on the symmetry breaking pattern that is seen in continuum
QCD but have to use its lattice version instead.

Our starting point is a partition function of the form \cite{ourSU2}
\begin{equation}
  \label{chPT}
\begin{split}
  \ln Z(m_v,m_s) \propto{} &VS(m_v,m_s) \\
   &-\dfrac12\dsum_Q K_Q \dsum_p \ln\bigl[\hat
       p^2+m_Q^2(m_v,m_s)\bigr] .
\end{split}
\end{equation}
$S(m_v,m_s)$ is the saddle-point contribution leading to a smooth
background, and the double sum represents the one-loop contribution due
to light composite (Goldstone) bosons.  The sum runs over the allowed
lattice momenta $p_\mu$ [$p_\mu = 2\pi n_\mu/L$ with integer $n_\mu$
and $\hat p^2\equiv 2\sum_\mu (1-\cos p_\mu)$] as well as over light
particles of type $Q$ with multiplicity $K_Q$ and mass $m_Q$.

In order to calculate the values for $K_Q$ and $m_Q$ we need to know
the pattern of chiral symmetry breaking which is given by
\begin{equation}
\begin{split}
  \SU(N_v&+N_s)\otimes\U(1)\otimes\SU(N_v+N_s)\otimes\U(1) \\
  &\to\SU(N_v+N_s)\otimes\U(1)
\end{split}
\end{equation}
for staggered fermions with three colors. Note that in this case the
$\U(1)$ symmetry is broken without an anomaly such that the number of
Goldstone bosons equals $(N_v+N_s)^2$. The masses of the flavor
non-diagonal mesons $\bar q_iq_j$ ($i\neq j$) are simply given by
$m^2=A(m_i+m_j)/2$.

\begin{figure}[b]
  \begin{center}
    \vspace*{-4mm}
    \epsfig{file=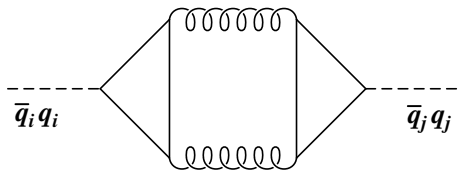,width=6cm}
  \end{center}
  \caption{Annihilation diagram.}
  \label{anni}
\end{figure}

The case of the flavor diagonal mesons is more difficult because we
must also consider annihilation according to Fig~\ref{anni}.
Since on the lattice with staggered fermions the broken $\U(1)$
symmetry is anomaly free the amplitude of this diagram is proportional
to $m_im_j$. This is in contrast to the continuum case where the
contribution of the diagram remains non-zero even in the chiral limit
because of the chiral anomaly.  For the mass-squared matrix $M^2$ of
these states $(\bar v_1v_1,\dots,\bar v_{N_v}v_{N_v},\bar
s_1s_1,\dots,\bar s_{N_s}s_{N_s})^T$ we get in addition to the usual
linear terms a contribution that is quadratic in the quark masses,
\begin{equation} \label{zterm}
\begin{split}
  M^2 ={} &A\diag(m_v,\dots,m_v,m_s,\dots,m_s) \\
   &+z\begin{pmatrix}
             m_v^2  & \cdots & m_v^2 & m_vm_s & \cdots & m_vm_s \\
             \vdots & \ddots & \vdots& \vdots & \ddots & \vdots \\
             m_v^2  & \cdots & m_v^2 & m_vm_s & \cdots & m_vm_s \\
             m_sm_v & \cdots & m_sm_v& m_s^2  & \cdots & m_s^2  \\
             \vdots & \ddots & \vdots& \vdots & \ddots & \vdots \\
             m_sm_v & \cdots & m_sm_v& m_s^2  & \cdots & m_s^2
            \end{pmatrix}
\end{split}
\end{equation}
where we have introduced an additional parameter $z$. In contrast to
our earlier publication \cite{ourSU2} for $\SU(2)$ we cannot neglect
this higher-order term here because it will turn out to be the leading
order term for $\chiconn$.

After diagonalization of $M^2$ we obtain five different eigenvalues:
$Am_v$, $Am_s$, $A(m_v+m_s)/2$ and $\lambda_\pm$ with multiplicities
given in Table~\ref{mtable}. The eigenvalues $\lambda_\pm$ are given
by
\begin{equation}
  \begin{split}
    \lambda_\pm ={}&\frac A2(m_v+m_s) + \frac z2(N_vm_v^2+N_sm_s^2) \\
    &\pm\frac12\bigl[A^2(m_v-m_s)^2+z^2(N_vm_v^2+N_sm_s^2)^2 \\
    &\qquad+2Az(m_v-m_s)(N_vm_v^2-N_sm_s^2)\bigr]^{1/2} \;.
  \end{split}
\end{equation}
\begin{table}[hbtp]
  \begin{tabular}{cc}
    \phantom{WWWWW}$m^2$\phantom{WWWWW} & 
    \phantom{WWWW}Multiplicity\phantom{WWWW} \\ 
    \hline\\[-2mm]
    $Am_v$      &    $N_v^2-1$      \\
    $Am_s$      &    $N_s^2-1$      \\
    $A(m_v+m_s)/2$ & $2N_vN_s$    \\
    $\lambda_-$ &    $1$            \\
    $\lambda_+$ &    $1$ \\[0mm]
  \end{tabular}
  \vspace*{2mm}
  \caption{The light particle spectrum for the gauge group
    $\SU(3)$.}
  \label{mtable}
\end{table}

Using Eq.~\eqref{chPT} and the multiplicities of Table~\ref{mtable} in
Eqs.~\eqref{a}--\eqref{c} we obtain
\begin{equation}
  \begin{split}
    \label{sigma}
    \lefteqn{\sigma(m_v,m_s) = C_0+C_cm_v+C_dN_sm_s } \\[0.5ex]
    &\qquad-\frac{AN_s}{L^4}\sum_p\frac1{2\hat p^2+A(m_v+m_s)}\\[0.5ex]
    &\qquad-\frac{m_vz}{2L^4}\sum_p
    \frac{(\hat p^2+Am_s)(2\hat p^2+Am_v)}
    {(\hat p^2+Am_v)^2(\hat p^2+Am_s+N_szm_s^2)}\:,
    \raisetag{4\baselineskip}
  \end{split}
\end{equation}
\begin{equation}
  \begin{split}
    \lefteqn{\chiconn(m) = C_c + \frac{A^2N_s}{4L^4}\sum_p
      \frac1{(\hat p^2+Am)^2} } \\
    &\qquad-\frac z{L^4}\sum_p \frac{(\hat p^2)^2}{(\hat p^2+Am)^2
      (\hat p^2+Am+N_szm^2)}\:,
  \end{split}
\end{equation}
\begin{equation}
  \begin{split}
    \lefteqn{\chidisc(m) = C_d +
      \frac{A^2}{4L^4}\sum_p\frac1{(\hat p^2+Am)^2}} \\
    &\qquad+\frac{z^2m^2}{2L^4}\sum_p \frac{(2\hat p^2+Am)^2}
    {(\hat p^2+Am)^2(\hat p^2+Am+N_szm^2)^2}\:.
    \raisetag{3\baselineskip}
  \end{split}
\end{equation}
In these expressions $C_0$, $C_c$, and $C_d$ (assumed to be constants)
describe the smooth background contributions.  For our quenched
simulations ($N_s=0$) the leading $m$-dependent term in the connected
scalar susceptibility is proportional to $z$ and thus to the
annihilation diagram.  This has the consequence that for $\chiconn$
the $m\to 0$ limit of chPT does not coincide with the $m\to\infty$
limit of chRMT anymore.  Therefore there is no mass range where both
theories coincide.

\begin{figure}[htbp]
  \begin{center}
   \epsfig{file=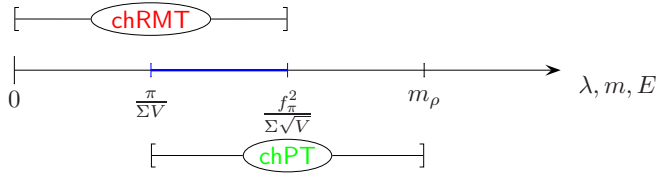,width=0.48\textwidth}\vspace*{3ex}
    \caption{Energy ranges for the applicability of chiral random
      matrix theory (chRMT) and chiral perturbation theory (chPT).}
    \label{ranges}
  \end{center}
\end{figure}

On the other hand, in the $\SU(2)$ case and for $\chidisc$ we can
identify three mass ranges.  Above $m\propto 1/\sqrt V$ chRMT fails
since the kinetic terms in the chiral Lagrangian become important.
Below $m\propto 1/V$ chPT becomes invalid because it does not include
the non-perturbative contributions to the partition function due to
the zero-momentum modes.  For sufficiently large volumes there is an
overlap region of chRMT and chPT which is shown in Fig.~\ref{ranges}.

In the thermodynamic limit ($L\to\infty$) and for $N_s=0$ our
observables are given by
\begin{align}
  \begin{split}
  \sigma(m) &= C_0 +(C_c-za_0)m -\frac{3Az}{32\pi^2}m^2\ln(Am) \\
     &\quad+{\cal O}(m^2) \:,
  \end{split}\\
  \chiconn(m) &= C_c-za_0 -\frac{3Az}{16\pi^2}m\ln(Am) +{\cal
                 O}(m)\:,\\
  \begin{split}
  \chidisc(m) &= {-\frac{A^2}{64\pi^2}\ln(Am)} +C_d -\frac{A^2a_1}4
     -\frac{A^2}{64\pi^2} \\
      &\quad+{\cal O}(m\ln m) \;.
  \end{split}
\end{align}
The numerical constants $a_0$ and $a_1$ take the values $a_0 = 0.1549$
and $a_1 = -0.03035$.

It is interesting to compare these formulae with the corresponding
results for quenched $\SU(2)$ (see Ref.~\cite{ourSU2}).  In $\SU(2)$
the leading $m$-dependent term in $\sigma$ goes with $m\ln(Am)$, while
in quenched $\SU(3)$ this term is absent.  Similarly we see that in
$\SU(2)$ the leading term in $\chiconn$ goes with $\ln(Am)$, which is
again absent here.  So chiral perturbation theory also predicts that
in quenched $\SU(3)$ the generic leading term, found in $\SU(2)$ and
dynamical $\SU(3)$, is missing.  This is the same sort of result that
we saw in chRMT in Eq.~\eqref{lead_term}.

\begin{figure}[htbp]
  \vspace*{-4mm}
  \begin{center}
    \epsfig{file=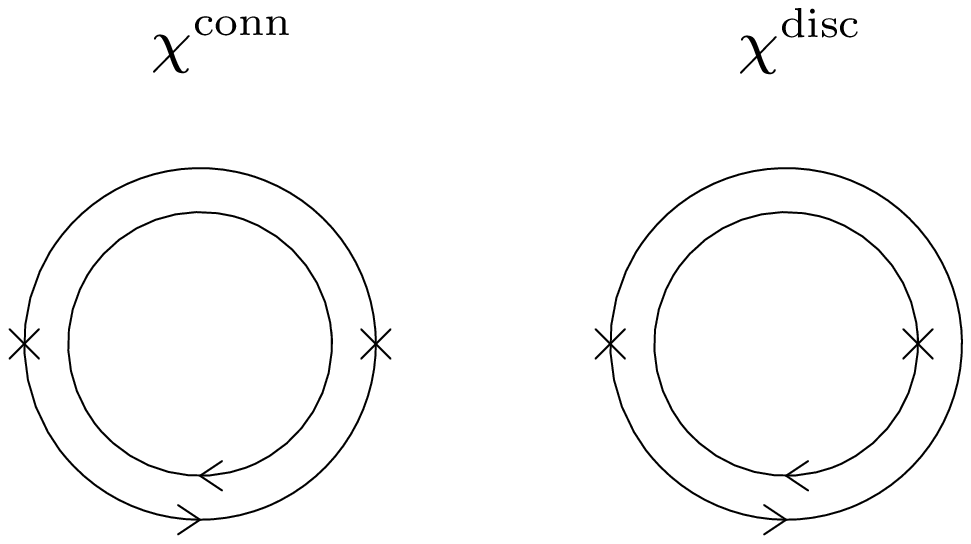,width=0.45\textwidth}
  \end{center}
  \caption{Goldstone meson contribution to $\chiconn$ and
    $\chidisc$.}
  \label{new_fig_a}
\end{figure}

\begin{figure}[htbp]
  \vspace*{-8mm}
  \begin{center}
    \epsfig{file=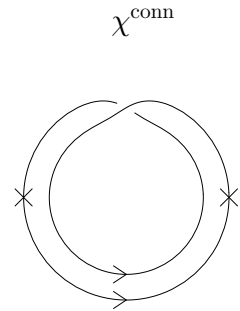,width=0.165\textwidth}
  \end{center}
  \caption{Goldstone baryon contribution to $\chiconn$ for gauge
    group $\SU(2)$.}
  \label{new_fig_b}
\end{figure}

What is the physical reason for this difference?  Chiral perturbation
theory lets us understand the cause.  In Fig.~\ref{new_fig_a} we show
the simplest Goldstone meson contributions to $\chiconn$ and
$\chidisc$.  The $\times$ represents a $\bar\psi\psi$ operator.  In
$\chiconn$ both operators are on the same quark line, in $\chidisc$
they are always on different quark lines (this is the reason for the
nomenclature).  We see that the diagram for $\chidisc$ has no
spectator quark loops, so it survives when we take the quenched limit.
The diagram for $\chiconn$ has a spectator loop, and so it vanishes in
the quenched approximation, when all spectator loops are ignored.  Why
then does $\chiconn$ for quenched $\SU(2)$ still have a logarithmic
term?  This is because, as emphasized in \cite{ourSU2}, $\SU(2)$ has
``Goldstone baryons'' as well as Goldstone mesons.  These are
two-quark states with masses that vanish in the chiral limit, just
like the more familiar Goldstone bosons of the other $\SU(N_c)$
groups.  In Fig.~\ref{new_fig_b} we sketch a Goldstone baryon
contribution to $\chiconn$ which survives in the quenched
approximation.
\\[1mm]

\section{Comparison with lattice data}

We want to confront the lattice data with the predictions from chPT
just described concentrating solely on the disconnected chiral
susceptibility $\chidisc(m)$ because for the other two quantities
there is no common range of applicability of chRMT and chPT.  Since
the terms proportional to $z$ are of higher order we will neglect them
in the following.  It is important to note that besides the Goldstone
boson $\pi$ the meson spectrum contains also $15$ would-be Goldstone
bosons whose masses do not vanish in the chiral limit $m\to0$.
(Remember that one generation of staggered fermions corresponds to
four flavors in the continuum limit.)  Instead we expect their masses
$m_i$ to depend on $m$ according to $m_i^2=A_im+B_i$.  They all
contribute a term of the form $\frac{A_i^2}{4L^4}\sum_p(\hat
p^2+A_im+B_i)^{-2}$.  Besides the Goldstone boson $\pi$ (with
$B_\pi=0$) the theory would allow for up to $7$ different boson masses
\cite{Golterman}.  Since we cannot afford so many fit parameters we
introduce just a single ``effective'' $B$ for the $15$ would-be
Goldstone bosons and set $A_i=A$.  The susceptibility $\chidisc$ then
becomes
\begin{equation}
\begin{split}
  \label{chidiscB}
  \chidisc(m) = C_d + \frac{A^2}{4L^4}&\sum_p\biggl[\frac1{(\hat
    p^2+Am)^2} \\
  &+ 15\,\frac1{(\hat p^2+Am+B)^2}\biggr]\;.
\end{split}
\end{equation}

The parameters $A$, $B$, and $C_d$ have been fitted jointly for all
lattice sizes that are available for our particular values of $\beta$.
The parameters are, in principle, functions only of $\beta$ but not of
$L$.  The fit interval has been chosen such that its left border is
within the overlap region of chRMT and chPT where the data show the
asymptotic behavior of both theories.  The large-$m$ cutoff is more
difficult, because one has to be careful not to extend the fit into
regions where our version of chPT is not applicable anymore.  We
therefore extended the interval until we found a stable plateau of the
parameter $A$.

\begin{table}[t]
    \begin{tabular}{cr@{.}lclc}
      $\beta$ & \Cc{$A$}   & $B$ & \cc{$C_d$} & $f_\pi$ \\
      \hline\\[-3mm]
      $5.2$ & $ 5$&$9(9) $ & $1.5(16)$ & $-1.6(2) $ & $0.30(3)$ \\
      $5.4$ & $ 7$&$2(9) $ & $0.55(4)$ & $-3.4(12)$ & $0.25(2)$ \\
      $5.6$ & $10$&$0(18)$ & $0.29(5)$ & $-7.0(33)$ & $0.15(2)$ \\
      $5.7$ & $ 7$&$8(15)$ & $0.11(4)$ & $-4.2(21)$ & $0.13(3)$
    \end{tabular}
    \vspace*{2mm}
    \caption{The fit parameters $A$, $B$, and $C_d$ as well as the
      values of $f_\pi$ obtained from Eq.~\protect\eqref{fpi}.}
    \label{results}
\end{table}

\begin{figure}[ht]
  \begin{center}\vspace*{-0.9ex}
    \epsfig{file=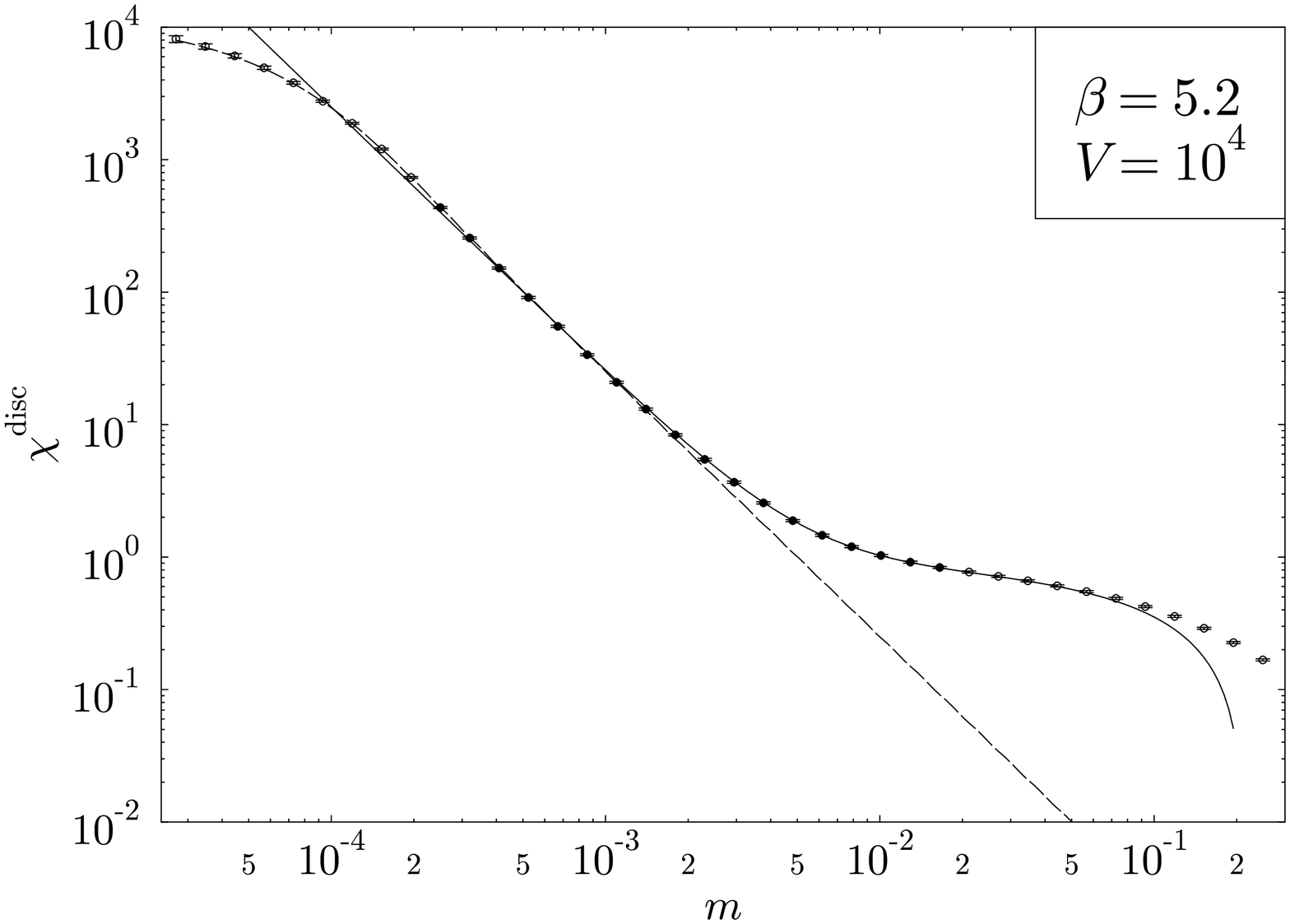,width=76.5mm}\\[2mm]
    \epsfig{file=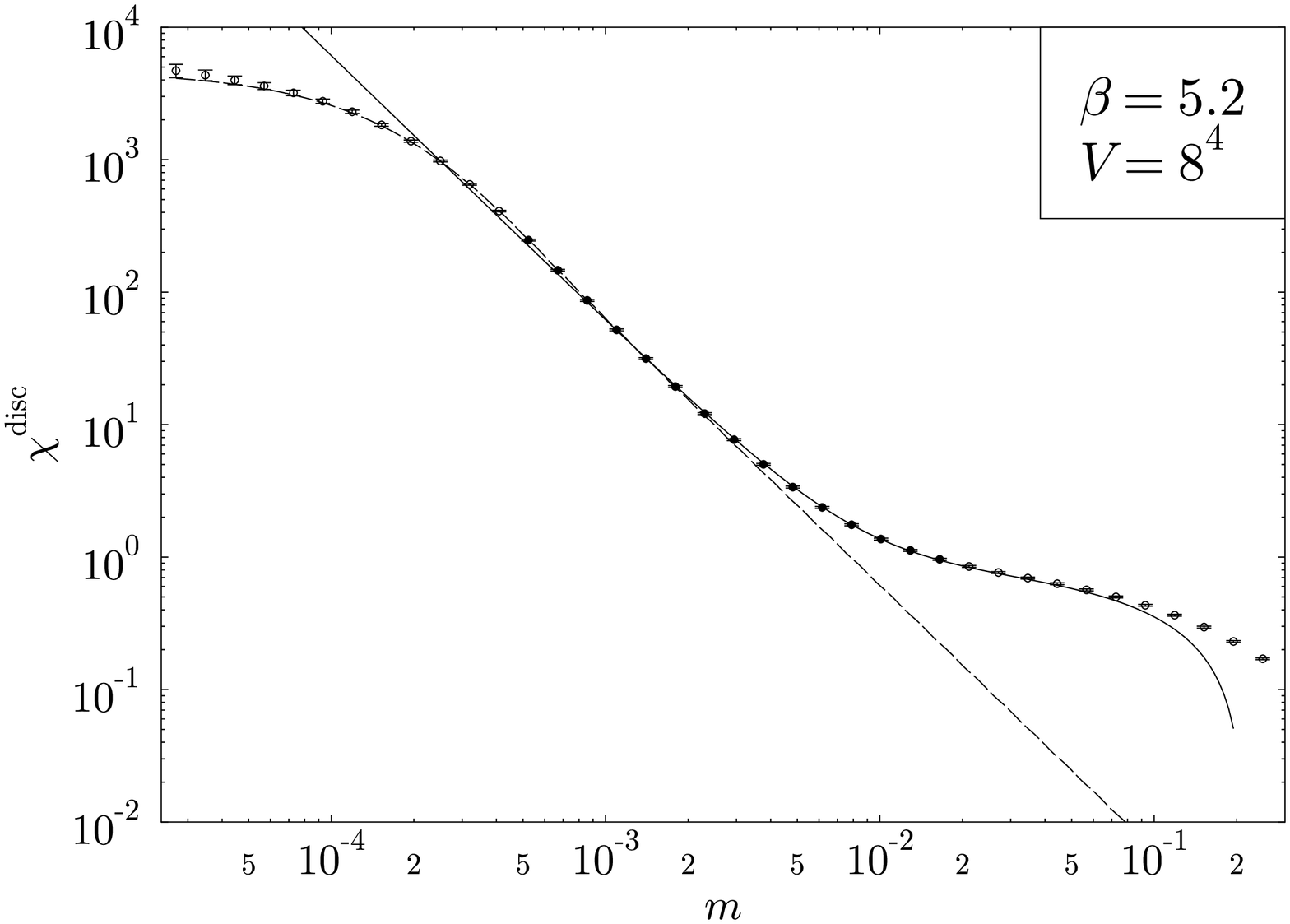,width=76.5mm}\\[2mm]
    \epsfig{file=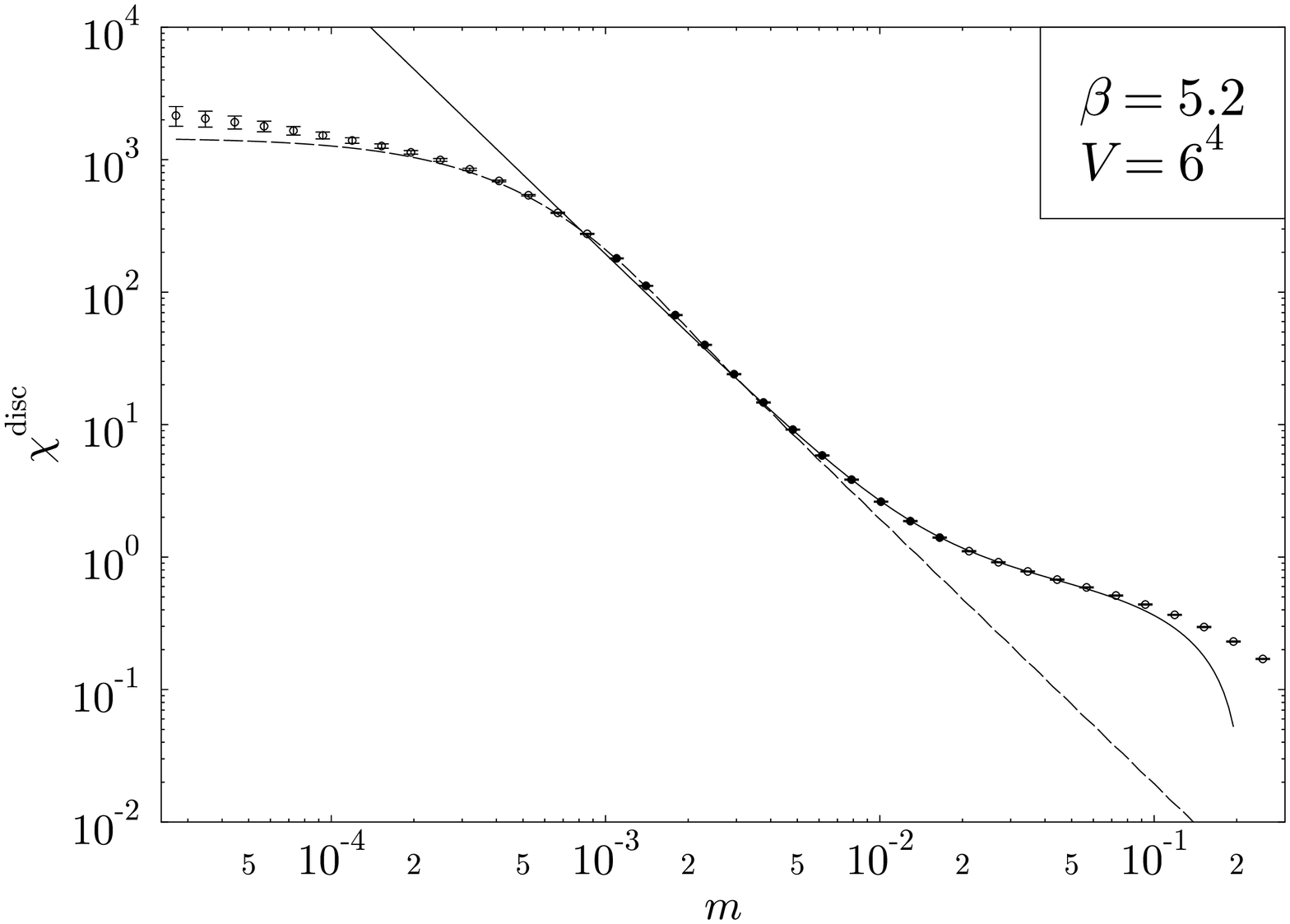,width=76.5mm}\\[4mm]
    \caption{The chiral susceptibility $\chidisc$ at $\beta=5.2$ for
      $L=6,8,10$ plotted against $m$.  The dashed curve represents
      the chRMT prediction.  The full curve results from a joint fit
      of all lattice volumes with the chPT formula
      \protect\eqref{chidiscB}.  The fit parameters are given in
      Table~\protect\ref{results}.}
    \label{5_2}
  \end{center}
\end{figure}
\begin{figure}[ht]
  \begin{center}
    \epsfig{file=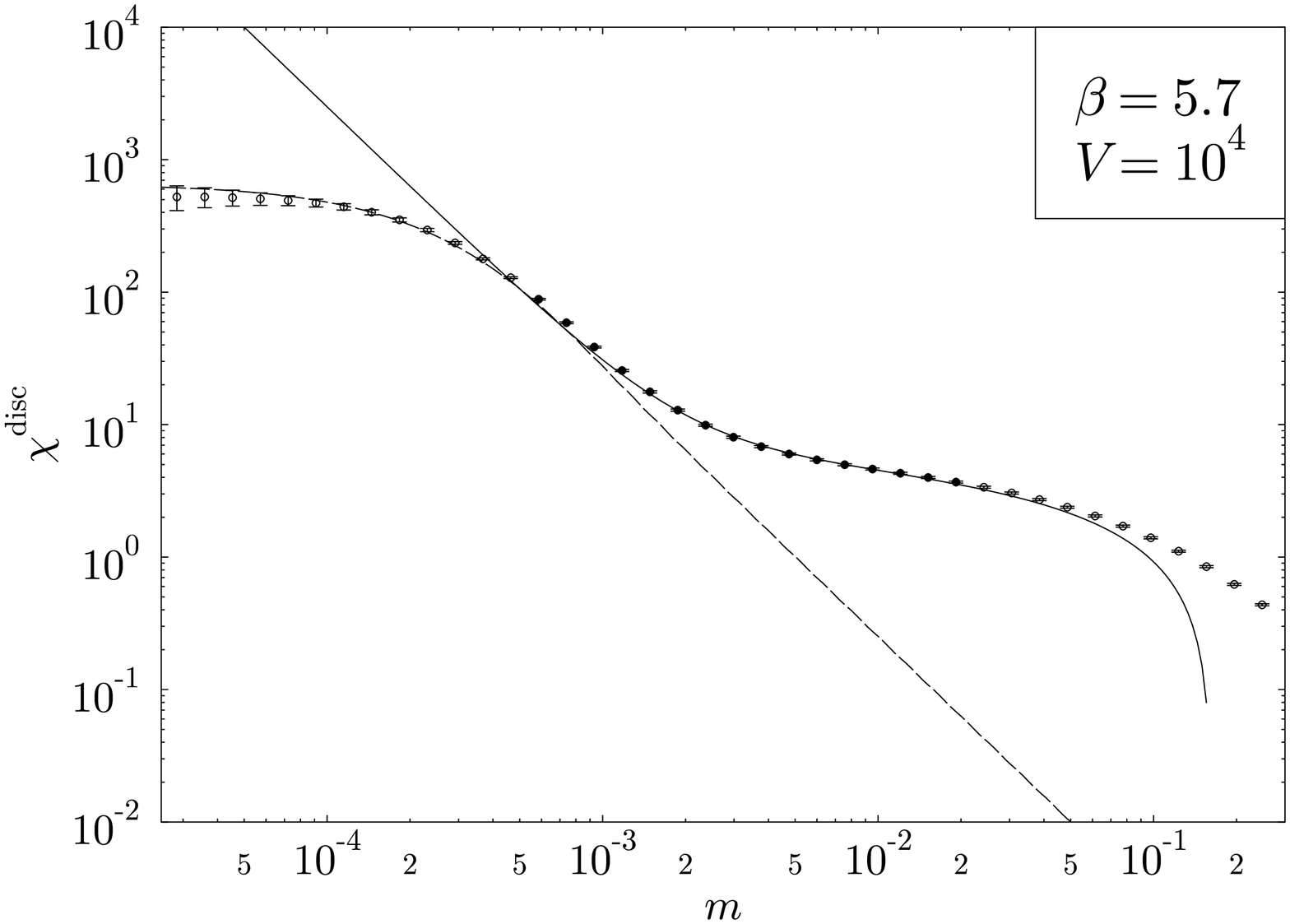,width=75mm}\\[2mm]
    \epsfig{file=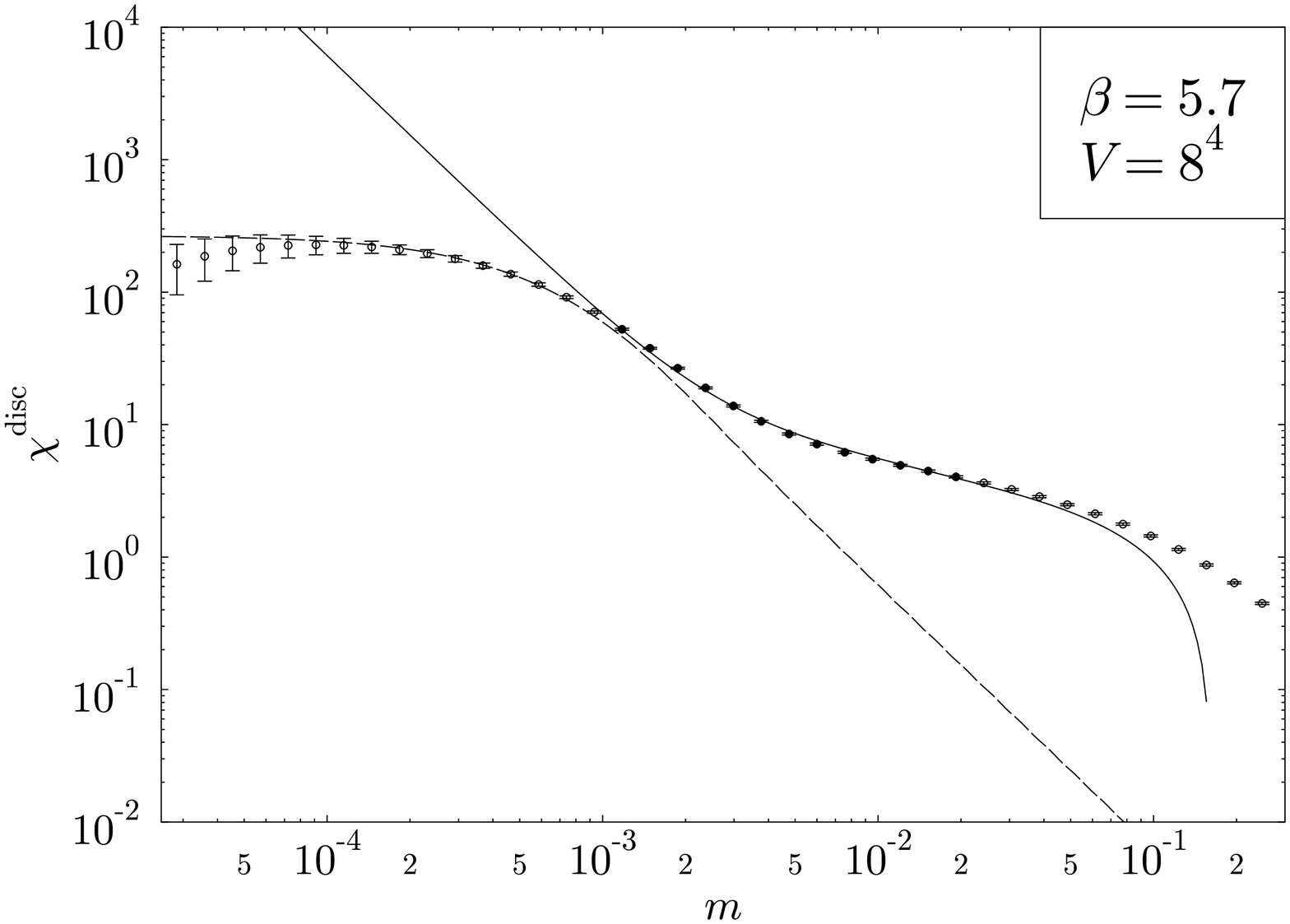,width=75mm}\\[4mm]
    \caption{The chiral susceptibility $\chidisc$ at $\beta=5.7$ for
      $L=8,10$ plotted against $m$.  The meaning of the curves is the
      same as in Fig.~\protect\ref{5_2}.  The fit parameters are given
      in Table~\protect\ref{results}.}
    \label{5_7}
  \end{center}
\end{figure}
\begin{figure}[ht]
  \begin{center}
    \epsfig{file=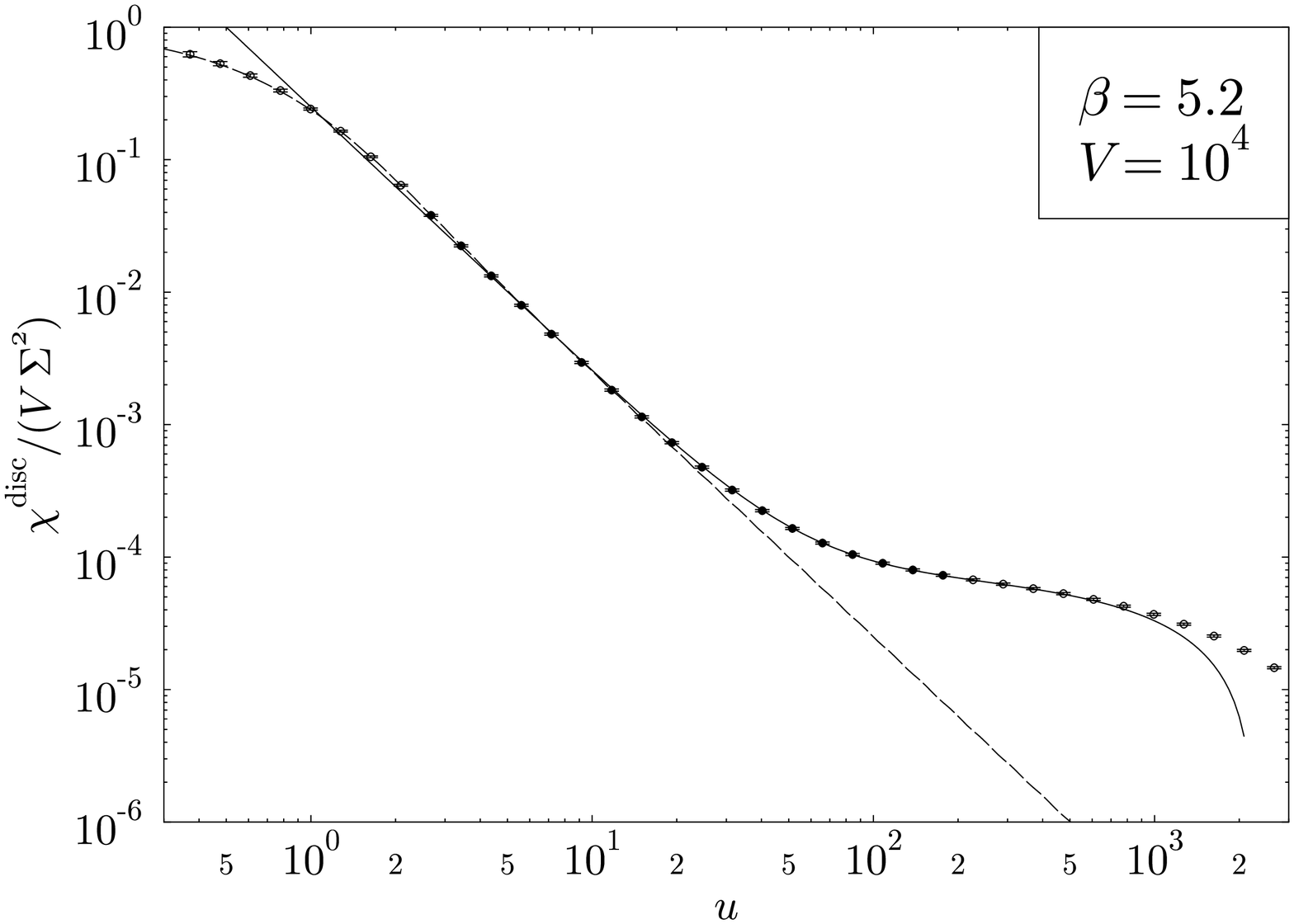,width=73mm}\\[1mm]
    \epsfig{file=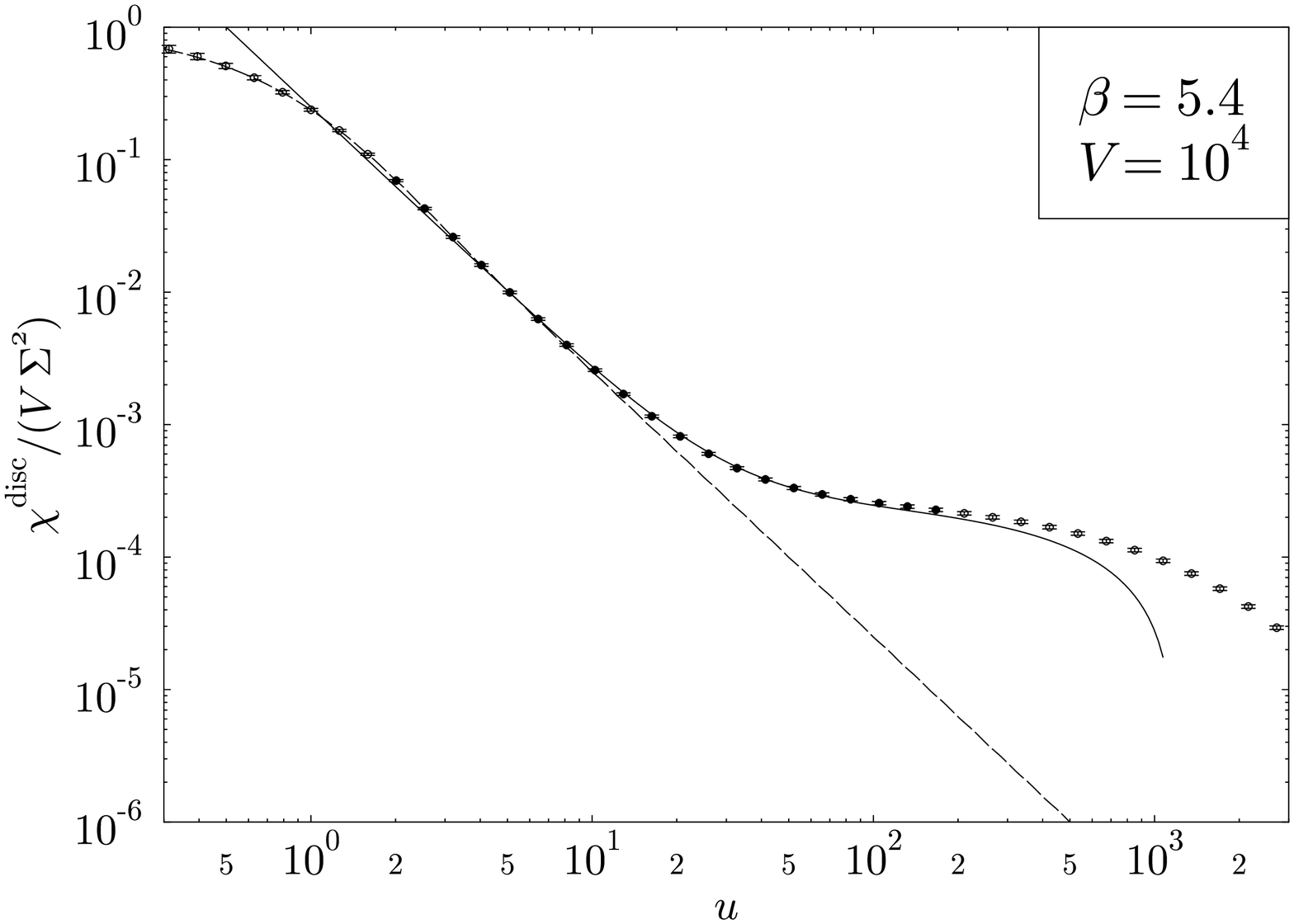,width=73mm}\\[1mm]
    \epsfig{file=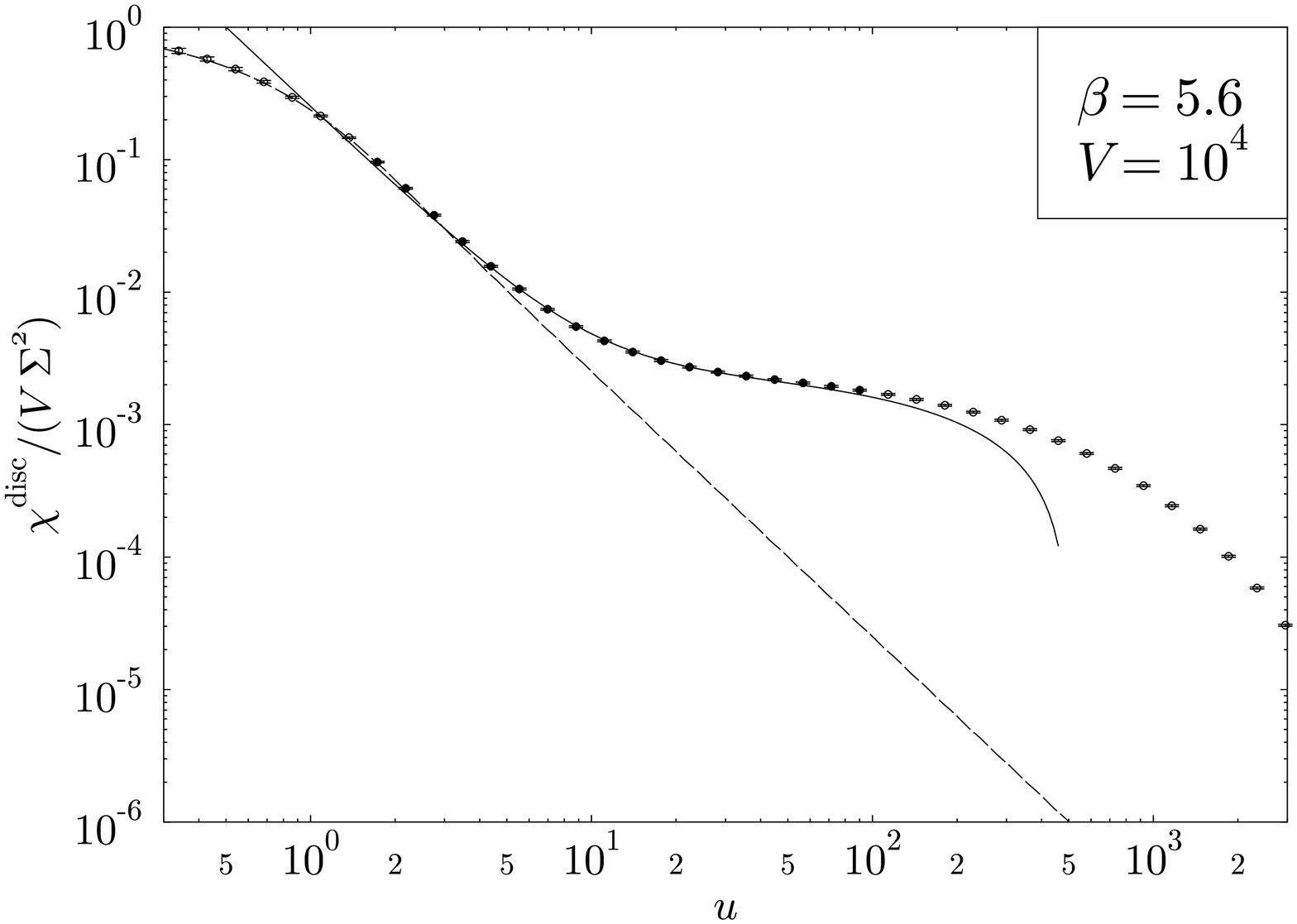,width=73mm}\\[1mm]
    \epsfig{file=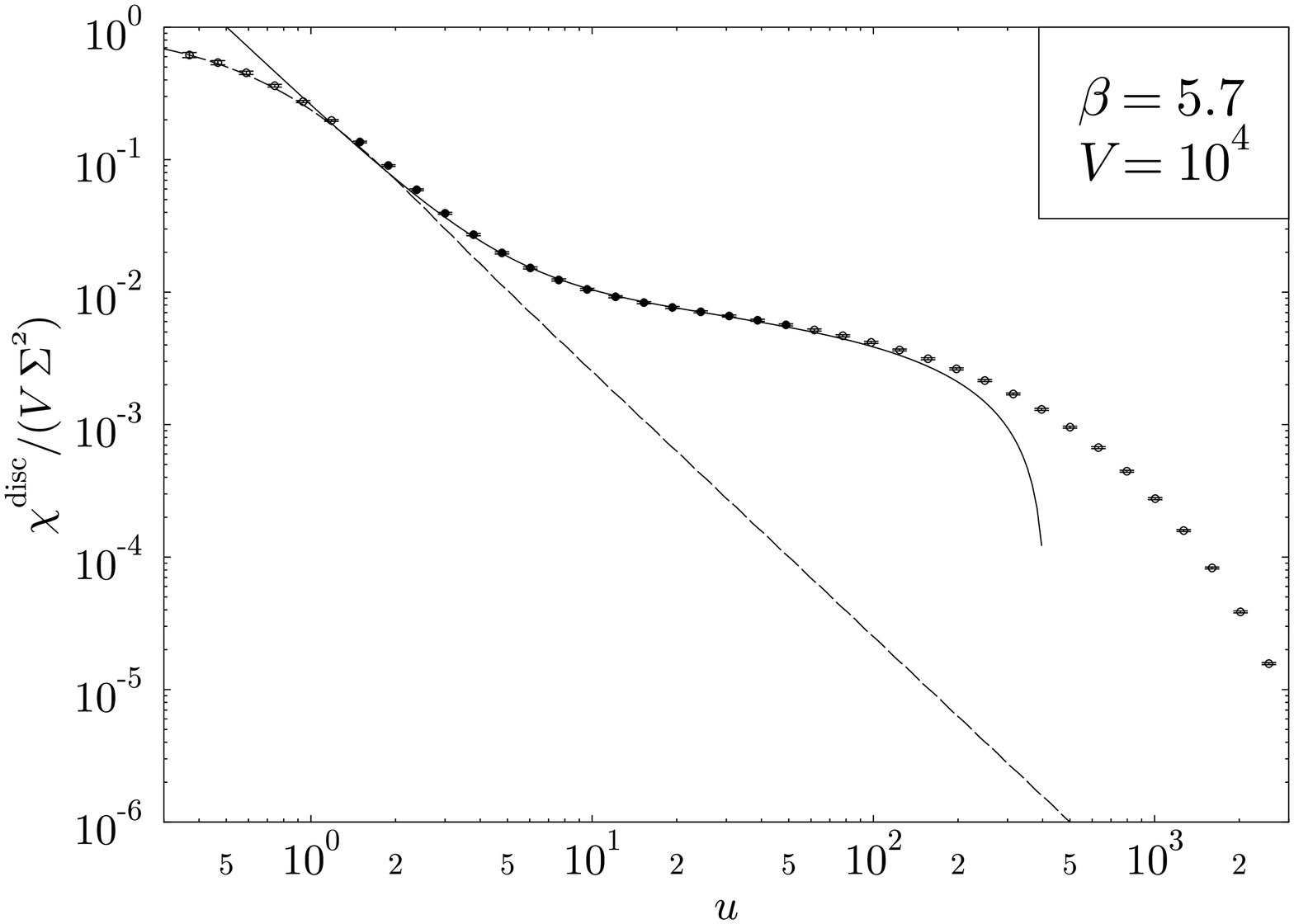,width=73mm}\\[3mm]
    \caption{The chiral susceptibility $\chidisc/(V\Sigma^2)$ for
      $L=10$ and $\beta=5.2,5.4,5.6,5.7$ plotted against $u=mV\Sigma$.
      The meaning of the curves is the same as in
      Fig.~\protect\ref{5_2}.  The fit parameters are given in
      Table~\protect\ref{results}.}
    \label{u}
  \end{center}
\end{figure}

The results for the fit parameters are given in Table~\ref{results}.
One observes a monotonic decrease of the parameter $B$ as $\beta$ gets
larger.  This shows that the would-be Goldstone bosons become more and
more important when one approaches the continuum limit.  Such a
behavior is of course expected because in the continuum limit all $16$
(would-be) Goldstone bosons should have the same mass, at least as far
as quenching artifacts can be neglected.  The parameter $A$, on the
other hand, shows some indication of non-monotonicity.  This is also
not completely unexpected: Strong coupling calculations
\cite{Jolicoeur} show that $m_\pi^2/m$ has a value of about $4.6$ at
$\beta=0$ and increases as $\beta$ grows.  But for $\beta\to\infty$
the ratio $m_\pi^2/m$ having dimension of a mass has to decrease to
zero in accordance with the renormalization group.  Although the
strong coupling expansions involve dynamical fermions, one would not
be surprised to find a similar behavior in quenched simulations.

In Figs.~\ref{5_2} and \ref{5_7} we show our data for $\chidisc$
together with the chRMT prediction and the chPT fits at $\beta=5.2$
and $5.7$, respectively.  This double logarithmic plot against $m$
confirms the expectation (see Fig.~\ref{ranges}) that the range (in
$m$) of common applicability of chRMT and chPT increases with $L$.
Indeed we read off from Fig.~\ref{ranges} that the ratio of the upper
end of this range over the lower end is roughly given by
$f_\pi^2L^2/\pi$.  In Fig.~\ref{u} we compare the $L=10$ data for our
different $\beta$ values.  We plot $\chidisc/(V\Sigma^2)$ against
$u=mV\Sigma$ together with the chRMT prediction and the chPT fits.
The range of common applicability of chRMT and chPT decreases as
$\beta$ grows, because $f_\pi$ (in lattice units) must tend to zero in
the continuum limit.

The link with usual chPT is provided by the Gell-Mann--Oakes--Renner
relation that relates the parameter $A$ to $\Sigma$ and the pion decay
constant $f_\pi$,
\begin{equation}
  \label{fpi}
  A = \frac 2{f_\pi^2}\frac\Sigma4\;.
\end{equation}\clearpage
The (unrenormalized) values of $f_\pi$ calculated from this equation
are shown in Table~\ref{results}.

In Ref.~\cite{fpi} Goldstone boson masses $m_\pi$ have been computed
with quenched staggered fermions on a $16^3\times32$ lattice at
$\beta=5.7$.  Fitting these masses with the relation $m_\pi^2=A_\pi
m+B_\pi$ the authors find $A_\pi=7.96(5)$ and $B_\pi=0.004(1)$.  Our
value for $A$ at $\beta=5.7$ agrees very well with this result giving
us confidence that our chPT model captures the essential features of
the underlying physics.

\section{Summary}

In this paper we have studied spectral properties of the staggered
Dirac operator in quenched SU(3) lattice gauge theory in the phase
where chiral symmetry is spontaneously broken. From complete spectra
of the Dirac operator we have computed the scalar susceptibilities
$\chidisc$ and $\chiconn$ as functions of the (valence) quark mass
$m$. For small masses the low-lying eigenvalues of the Dirac operator
give the dominant contribution and the mass dependence follows the
predictions of chiral random matrix theory. This agreement holds for
masses which are smaller than the so-called Thouless energy.  In the
generic case, the Thouless energy scales like $L^2$, where $L$ is the
linear size of the lattice.  In physical terms, this behavior results
from the fact that the Compton wavelength of the lightest particles in
the theory, the Goldstone bosons, exceeds $L$ as long as $m$ lies
below the Thouless energy, and the susceptibilities are therefore
insensitive to the details of the dynamics.  For the gauge group SU(2)
the expected scaling behavior has been confirmed previously for both
susceptibilities \cite{Ber2,ourSU2}.  In the present case of gauge
group SU(3) we observed scaling with $L^2$ for $\chidisc$ \cite{our},
whereas for $\chiconn$ the Thouless energy was found to scale with
$L^{4/3}$. This exceptional behavior is explained as a quenching
artifact.

Above the Thouless energy the Goldstone bosons begin to fit into the
lattice volume and one enters the realm of (quenched) chiral
perturbation theory. The application of chiral perturbation theory to
our case is complicated by the subtle chiral properties of staggered
fermions.  In particular, it turns out that for our simulation
parameters the contributions from the would-be Goldstone bos\-ons
cannot be neglected.  Taking them into account by means of a rough
model we obtain a satisfactory description of our data, which also
allows us to determine the pion decay constant $f_\pi$. At our largest
$\beta$ value ($\beta = 5.7$) we could compare our result for $f_\pi$
with numbers from the literature and found nice agreement.

After completion of our work, a preprint \cite{Damgaard} appeared that
discusses related issues using (partially) quenched chiral
perturbation theory, also in a finite volume but already in the
continuum limit.

\acknowledgments

This work was supported by DFG, BMBF, DOE contracts DE-FG02-91ER40608
and DE-AC02-98CH10886, and the RIKEN-BNL Research Center.


\begin{thebibliography}{99}
\bibitem{ShuVer} E.V. Shuryak and J.J.M. Verbaarschot, Nucl.\
  Phys.\ {\bf A560}, 306 (1993).
\bibitem{review1} For a recent review on random matrix theory in
  general, see T. Guhr, A. M{\"u}ller-Groeling, and
  H.A. Weidenm{\"u}ller, Phys.\ Rep.\ {\bf 299}, 189 (1998) and
  references therein.
\bibitem{review2} For recent reviews on chRMT and QCD Dirac spectra,
  see J.J.M. Verbaarschot and T. Wettig, Annu.\ Rev.\ Nucl.\ Part.\
  Sci.\ {\bf 50}, 343 (2000);
  J.J.M. Verbaarschot, hep-ph/9902394 and references therein.
\bibitem{HV} M.A. Halasz and J.J.M. Verbaarschot,
  Phys.\ Rev.\ Lett.\ {\bf 74}, 3920 (1995).
\bibitem{EdwHel} R.G. Edwards, U.M. Heller, and R. Narayanan, Phys.\
  Rev.\ D {\bf 60}, 077502 (1999).
\bibitem{Berb98a} M.E. Berbenni-Bitsch, S. Meyer, A. Sch{\"a}fer,
  J.J.M. Verbaarschot, and T. Wettig, Phys.\ Rev.\ Lett.\ {\bf 80},
  1146 (1998).
\bibitem{Ma98} J.-Z. Ma, T. Guhr, and T. Wettig, Eur.\ Phys.\ J. A
  {\bf 2}, 87 (1998). 
\bibitem{Berb98b} M.E. Berbenni-Bitsch, S. Meyer, and T. Wettig,
  Phys.\ Rev.\ D {\bf 58}, 071502 (1998).
\bibitem{SU3} P.H. Damgaard, U.M. Heller, and A. Krasnitz, Phys.\
  Lett.\ B {\bf 445}, 366 (1999).
\bibitem{our} M. G{\"o}ckeler, H. Hehl, P.E.L. Rakow,
  A. Sch{\"a}fer, and T. Wettig, Phys.\ Rev.\ D {\bf 59}, 094503
  (1999).
\bibitem{Jac96} J.J.M. Verbaarschot, Phys.\ Lett.\ B {\bf 368}, 137
  (1996).
\bibitem{Jac98} J.C. Osborn and J.J.M. Verbaarschot, Phys.\ Rev.\
  Lett.\ {\bf 81}, 268 (1998); Nucl.\ Phys.\ {\bf B525}, 738 (1998).
\bibitem{Zahed98} R.A. Janik, M.A. Nowak, G. Papp, and I. Zahed,
  Phys.\ Rev.\ Lett.\ {\bf 81}, 264 (1998).
\bibitem{Ber2} M.E. Berbenni-Bitsch, M. G{\"o}ckeler, T. Guhr, A.D.
  Jackson, J.-Z. Ma, S. Meyer, A. Sch{\"a}fer, H.A. Weidenm{\"u}ller,
  T. Wettig, and T. Wilke, Phys.\ Lett.\ B {\bf 438}, 14 (1998).
\bibitem{ourSU2} M.E. Berbenni-Bitsch, M. G{\"o}ckeler, H. Hehl,
  S. Meyer, P.E.L. Rakow, A. Sch{\"a}fer, and T. Wettig, Phys.\
  Lett.\ B {\bf 466}, 293 (1999).
\bibitem{Forrester} P.J. Forrester, Nucl.\ Phys.\ {\bf B402}, 709
  (1993).
\bibitem{beyond} M.E. Berbenni-Bitsch, M. G{\"o}ckeler, H. Hehl,
  S. Meyer, P.E.L. Rakow, A. Sch{\"a}fer, and T. Wettig, Nucl.\
  Phys.\ Proc.\ Suppl.\ {\bf 83--84}, 974 (2000).
\bibitem{topology} P.H. Damgaard, U.M. Heller, R. Niclasen, and
  K. Rummukainen, Phys.\ Rev.\ D {\bf 61}, 014501 (2000).
\bibitem{Golterman} M.F.L. Golterman, Nucl.\ Phys.\ {\bf B273}, 663
  (1986).
\bibitem{Jolicoeur}
  T. Jolic{\oe}ur, H. Kluberg-Stern, M. Lev, A. Morel, and
  B. Petersson, Nucl.\ Phys.\ {\bf B235}, 455 (1984).
\bibitem{fpi} R. Gupta, G. Guralnik, G.W. Kilcup, and S.R. Sharpe,
  Phys.\ Rev.\ D {\bf 43}, 2003 (1991).
\bibitem{Damgaard}
  P.H. Damgaard, hep-lat/0105010.
\end{thebibliography}
\end{document}